\documentclass[useAMS,usenatbib]{mn2e}

\usepackage{amssymb,amsmath}
\usepackage{graphicx}
\usepackage{color}
\title[Radiative cooling functions for primordial molecules]{Radiative cooling functions for primordial molecules}
\author[C. M. Coppola, L. Lodi, J. Tennyson]{C. M. Coppola$^{1,2}$\thanks{E-mail:
carla.coppola@chimica.uniba.it (CMC); l.lodi@ucl.ac.uk (LL); j.tennyson@ucl.ac.uk (JT)}, L. Lodi $^{1}$, J. Tennyson $^{1}$\\
$^{1}$Department of Physics and Astronomy, University College London, Gower Street, London WC1E 6BT\\
$^{2}$Universit\`{a} degli Studi di Bari, Dipartimento di Chimica, Via Orabona 4, I-70126, Bari, Italy}
\begin{document}

\date{Accepted 2011 March 15.  Received 2011 March 15; in original form 2011 February 10}

\pagerange{\pageref{firstpage}--\pageref{lastpage}} \pubyear{2002}

\maketitle

\label{firstpage}

\begin{abstract}
Cooling of primordial gas plays a crucial role in the birth of the first structures in our Universe. Due to the
low fractional abundance of molecular species at high redshifts, spontaneous emission rather than collisions represents the most efficient way to cool the
pristine plasma. In the present work, radiative
cooling functions are evaluated for the diatomic species HD, HD$^+$, HeH$^+$, LiH and LiH$^+$.
Cooling functions for the triatomic ions H$_3^+$ and H$_2$D$^+$ are also considered. Analytic fits as functions of
temperature are provided.
\end{abstract}


\section{Introduction}

In its present shape, our Universe appears to be a large ensemble of bound structures. These are the results of an
intricate net of chemical and physical processes of increasing complexity, the first of which is fragmentation
of the primordial gas clouds, see e.g. \cite{b15}.
One of the fundamental ingredient of this ensemble of processes
is the cooling of pristine plasma. Indeed, the birth of the first objects during the evolution of the primordial gas could not have
occured without significant cooling, as the gas temperatures involved in the
initial phase of the Universe are too high to allow the early clouds to gravitationally collapse. Among the
plasma constituents, molecules play the most significant role as coolers of the pristine gas.
Molecular cooling takes place mainly via rotational-vibrational (ro-vibrational) transitions between
the internal, nuclear motion degrees of freedom.
For temperatures below about 8000~K molecular cooling processes are much more efficient
than the corresponding atomic ones, see e.g. \cite{b18bis}.

In models, the standard way of capturing the effects of cooling by a given species is via the {\it cooling function}, which 
gives the energy lost per second at a specified temperature.
Cooling functions are made up by two main contributions. First, molecules spontaneously emit radiation while making
ro-vibrational transitions; second, collision induced emission can occur; this is particularly important for high symmetry species where spontaneous emission can be very weak.

In a plasma, two key features render a molecule an efficient cooler at low temperature: its fractional abundance and its dipole moment.
In the chemistry of the primordial Universe H$_2$ represents the most abundant molecular species. Its role
as an efficient primordial coolant has been widely described, e.g. in
the formation of the first stars (e.g. Abel et al.~2002) and galaxies (e.g. Bromm et al.~2009, Benson~2010).
Both collisional and radiative H$_2$ cooling contributions have been considered in the literature. The role  of
different collisional partners (H$_2$, He, H$^+$ and $e^-$) was considered by  Glover \& Abel~(2008).
Together with H$_2$, the cooling properties of its cation H$_2^+$ due both to radiative and collisional
 (H and $e^-$ impact excitation) pathways have been described (Suchkov \& Shchekinov~1978).
Both H$_2$ and H$_2^+$ are customarily included in molecular cooling models because of their high relative fractional abundance.
However, because of their lack of a permanent dipole moment,
only weak electric-quadrupole or magnetic-dipole transitions are allowed in these species, severly limiting their cooling efficiency, particularly at low density.

Considering the mole fractions of chemical species given by
chemical networks such as the one by Galli \& Palla (1998) (hereafter, GP98),
other molecular species should be added to the list of coolants.
Grouping the primordial molecules as deuterated or helium-containing or
lithium enriched, HD, HD$^+$, HeH$^+$, LiH and LiH$^+$ represent the candidates to focus on. 
Their fractional abundances are expected to have the following freeze-out values ($z\approx 10$):
$f$(HD)~$\simeq 10^{-9}$; $f$(HD$^+$)~$\simeq 10^{-18}$; $f$(HeH$^+$)~$\simeq 10^{-12}$; $f$(LiH)~$\simeq 10^{-19}$;
$f$(LiH)$^+$~$\simeq 10^{-17}$. It is hard to form triatomic molecules
at high $z$, but the most likely species are H$_3^+$ and H$_2$D$^+$;
their fractional
abundances are thought to be approximately $f$(H$_3^+$)~$\simeq 10^{-18}$ and
$f$(H$_2$D$^+$)$^+$~$\simeq 10^{-19}$, respectively.
Collisional contributions to the cooling
 have been evaluated in difLferent density regimes for 
HD ad LiH (\cite{b3}, GP98).
In particular, the
effect of the most abundant collision partner H on this cooling pathway has been investigated.
However, in order to complete this description, the contribution of radiative cooling has to be taken into account as well.

The present work is organized as follows. In Section~\ref{conti} the equations used to compute the radiative
 cooling function are presented, describing the methods adopted for each molecule. Some details on previous calculations on
collisional cooling functions are given. Section~\ref{conclusioni} summarizes our results and  provides fits to the calculated data.

\section[]{Radiative cooling function: definition and calculations}\label{conti}
The total cooling function for the chemical species X is usually approximated as (e.g. Hollenbach \& McKee 1979, GP98):
\begin{equation}
 \Lambda_\mathrm{X}=\Lambda_\mathrm{X,LTE}\left( 1+\frac{n^{\mathrm{cr}}}{n\mathrm{(collider)}}\right) ^{-1}
\label{generalcooling}
\end{equation}
where $\Lambda_\mathrm{X,LTE}$ represents the local thermodynamic equilibrium radiative contribution to the cooling function, $n^\mathrm{cr}$ and $n\mathrm{(collider)}$ are the critical
 and collider density, respectively. The former is defined by:
\begin{equation}
 n^{\mathrm{cr}}=\left(\frac{\Lambda_\mathrm{X,LTE}}{\Lambda_\mathrm{X}(n\mathrm{(collider)})}n\mathrm{(collider)}\right)_{n\mathrm{(collider)}\rightarrow 0}
\label{ncr}
\end{equation}
where $\Lambda_\mathrm{X}(n\mathrm{(collider)}\rightarrow 0)$ is the low-density limit of the cooling function.
Eq.~(\ref{generalcooling}) is valid for all density regimes of the collider particle.
In the high-density limit $\Lambda_\mathrm{X}$ reduces to $\Lambda_\mathrm{X,LTE}$,
implying that in this regime local thermal equilibrium is reached and that cooling is
dominated by ro-vibrational transitions of the molecular species X.

Denoting with the letters ``$u$'' and ``$l$'' repectively the upper and lower ro-vibrational
states between which a spontaneous transition may occur, the
radiative cooling function is defined as:
\begin{equation}
W(T)=\frac{1}{Z(T)}\sum_{u,l} A_{ul}(E_u-E_l)(2J_u-1) g_u n_u
\label{w}
\end{equation}
where $Z(T)$
represents the partition function of the molecule, $A_{ul}$ the Einstein coefficient
for the $u\rightarrow l$ transition, $J_u$ the rotational angular momentum 
of the upper rotational state, $g_u$ the nuclear spin degeneracy (equal to 1
 for all calculations below),
 $E_u$ and $E_l$ the energies of the upper and lower state, and $n_u$ the population of the upper level.
As discussed above, under LTE conditions it holds:
\begin{equation} \begin{split}\label{wlte}
W(T)  & \equiv \Lambda_\mathrm{LTE}(T)\\
n_u   & = e^{-E_u/(k_B T)}\\
Z(T)  & =\sum_{i} (2J_i+1) \, g_i \, e^{-E_i/(k_B T)}
\end{split}
\end{equation}
Radiative cooling functions have been computed in the present work using the equations above. 
As $k_B^{-1} = 1.43878$~cm~K, the Boltzmann factor $n_u$
implies that at a temperature of $1000$~K the cooling function is mostly determined by 
transitions with $E_u \lesssim 700$~cm$^{-1}$.

Data on the emission probabilities and energy levels of the molecular
species under analysis are therefore required in order to compute $\Lambda_{\mathrm{LTE}}$.
Depending on the available data, different strategies have been adopted for each molecular system considered.
For HD and HeH$^+$, the relevant data have been calculated respectively by Abgrall et al.~(1982) and Engel et al.~(2005).
To obtain the necessary data for the other diatomic molecules under investigation the program {\sc level} 8.0 by Le Roy (2007) was used.
This
program solves the radial one-dimensional Schr\"{o}dinger equation
\begin{equation}\begin{split}\label{EdS}
&-\frac{\hslash^2}{2\mu}\frac{\mathrm{d}^2\Psi(r)}{\mathrm{d}r^2}+V_J(r)\Psi(r)=E_{v,J}\Psi(r)\\
& V_J(r) =V(r) + \frac{\hbar^2 [J (J+1)-\Omega^2]}{2 \mu r^2}
\end{split}\end{equation}
where $\mu$ is the reduced mass of the system, $J$ the total rotational angular moment, $\Omega$
the projection of the electronic angular momentum along the internuclear axis ($\Omega=0$ for the states of our interest) and
$V(r)$ is the diatomic potential energy curve, given as input.
Once the ro-vibrational wavefunctions $\Psi_{\nu,J}$ have been obtained {\sc level} can also calculate
the Einstein coefficients by
\begin{equation}\begin{split}\label{formula.Einstein}
 & A_{\nu'J'\rightarrow \nu"J"} =  \\
 & \left( \frac{16 \pi^3}{3 \varepsilon_0 \hbar} \right) \frac{S(J',J'')}{2J'+1} \, \tilde{\nu}^3 \mid\langle\psi_{\nu',J'}|M(r)|\psi_{\nu",J"}\rangle\mid^2
\end{split}\end{equation}
 where 
$S(J',J'')$ are
the H\"{o}nl-London rotational intensity factors, resulting from integration over the rotational degrees of freedom,
$\tilde{\nu}$ is the wavenumber of the transition, $M(r)$ is the diatomic dipole moment curve and is given as input.
In the common case where $\tilde{\nu}$ is measured in cm$^{-1}$ and $M$ in Debye the constant in brackets in Eq.~(\ref{formula.Einstein}) assumes the value $3.1361891\times10^{-7}$~cm$^3$ D$^{-2}$ s$^{-1}$.

Calculations were carried out for the electronic ground state of the molecular systems under investigation.
The following subsections examine each molecule considered and give detailed information on the potential energy and
dipole moment curves used to perform the calculations. Comparisons with existing data are also provided where possible.
\subsection{HD}
The role of HD in cooling phenomena of primordial plasma has been widely examinated in several studies (Flower et al. 2000, Lipovka et al. 2005, Ripamonti 2007). The effect of HD cooling on the formation of massive primordial stars was analysed by Yoshida et al.~(2007) and McGreer \& Bryan~(2008).
Despite its lower fractional abundance,
HD is a more efficient coolant than molecular hydrogen H$_2$, especially for $T<2000~$K (GP98).
This is because HD molecules possess a small permanent dipole moment ($\approx 8.3\times 10^{-4}$ D, Abgrall et al.~1982)
due to the asymmetry of the nuclei; by contrast, in non-deuterated molecular hydrogen
only very weak quadrupole transitions are allowed.

\begin{figure}
\includegraphics[width=0.5\textwidth]{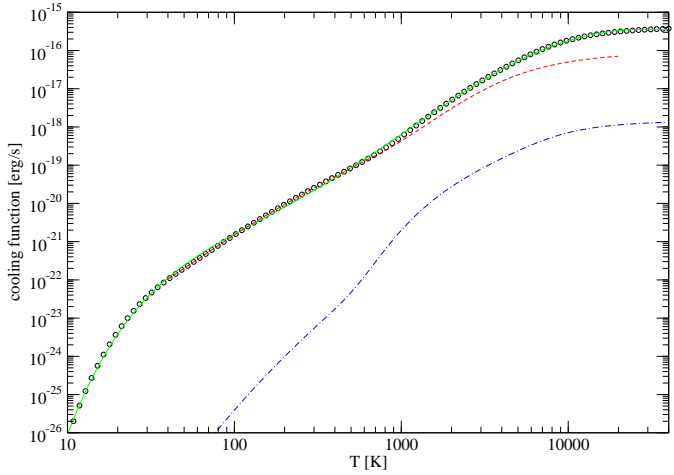}
\caption{HD radiative cooling function. {\it Circles}: dipole contribution, present calculation;
 {\it blue dotted-dashed curve}: quadrupole contribution;
{\it red dashed curve}: high-density limit fit by Lipovka et al. 2005;
{\it green solid curve}: fit given in Table~\ref{fits}.}
\label{HD}
\end{figure}

Lipovka et al. (2005) provide a two-parameter fit of the collisional cooling function, depending on temperature and
H fractional abundance. Here we evaluate
a radiative cooling function which takes into account both dipole and quadrupole electric transitions, for which we used the
calculations by Abgrall et al.~(1982).
Results are shown in Fig.~\ref{HD}, and compared with the high-density limit HD cooling function fit provided by Lipovka et al.~(2005),
where the contribution of the lowest ro-vibrational levels up to $v=3$ were considered.
A deviation from the high-density limit collisional contribution to cooling function is found for
temperatures above $700~$K, where highly excited ro-vibrational states are expected to be populated.

\subsection{HD$^+$}
The HD cation and its collisional contribution to the cooling during the evolution of primeval plasma
 has been examined by Glover \& Abel (2008), where calculations were
carried out assuming that there were no significant differences between the molecular ions H$_2^+$ and HD$^+$
 in the high-density cooling limit.
These authors provide an analytical expression for $\Lambda_{\mathrm{LTE}}$, calculated using
molecular data from Karr \& Hilico~(2006) on HD$^+$ (energy levels) and Peek et al.~(1979) on H$_2^+$ (quadrupole moments).

\begin{figure}
\includegraphics[width=0.5\textwidth]{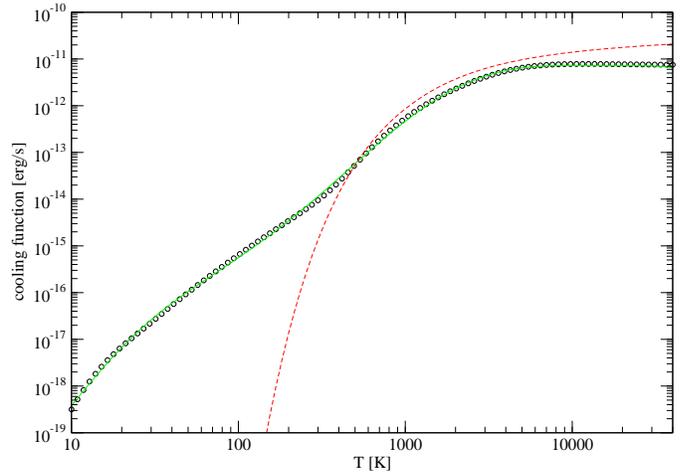}
\caption{HD$^+$ radiative cooling function: {\it Circles}: present calculation; {\it red dashed curve}: fit by Glover \& Abel 2008;
 {\it green solid curve}: fit given in Table~\ref{fits}.}
\label{HD+}
\end{figure}

Our calculations of the HD$^+$ radiative cooling function were carried out using the potential
 energy and dipole moment curves provided by Esry \& Sadeghpour~(1999), in which an adiabatic reformulation of the HD$^+$
Hamiltonian that recovers the isotopic splitting of electronic
states is presented. The resulting cooling curve is shown in Fig.~\ref{HD+},
and compared with the fit by Glover \& Abel~(2008). It should be noted that
our calculations predict significantly increased cooling at temperatures
below 100~K, where the cooling by HD$^+$ rotational transitions is very
much more efficient than the one due to its H$_2^+$ parent ion.


\subsection{HeH$^+$}
HeH$^+$ is the first molecule formed in the initial stages of chemical synthesis in the primordial Universe chemistry
 (GP98, Lepp et al.~2002, Dalgarno~2005, Hirata \& Padmanhaban 2006).
 It represents one of the reagents for the process:\\

\begin{figure}
\includegraphics[width=0.5\textwidth]{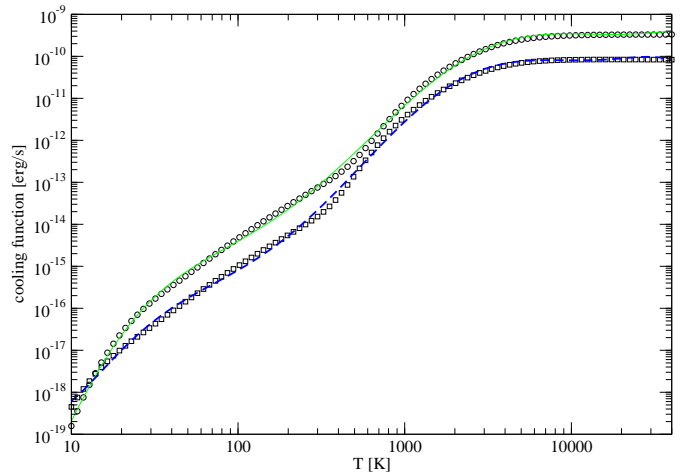}
\caption{HeH$^+$ radiative cooling function contributions. {\it Circles}: $^3$HeH$^+$; {\it squares}: $^3$HeD$^+$;
{\it green solid curve}: $^3$HeH$^+$ fit (Table~\ref{fits}); {\it blue dashed curve}: $^3$HeD$^+$ fit (Table~\ref{fits}).}
\label{HeH+_a}
\end{figure}

\begin{figure}
\includegraphics[width=0.5\textwidth]{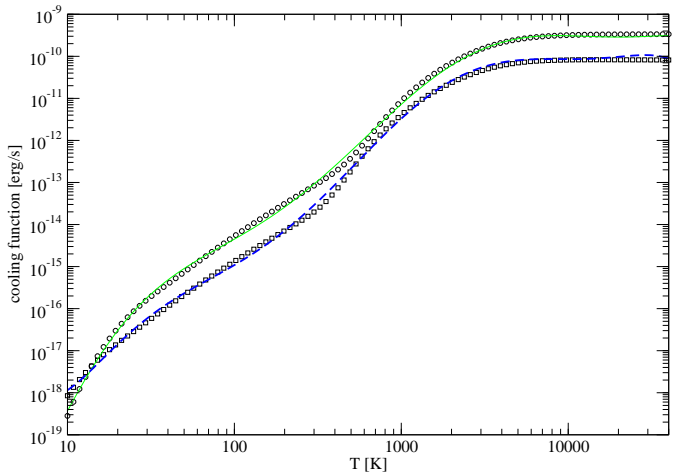}
\caption{HeH$^+$ radiative cooling function contributions. {\it Circles}: $^4$HeH$^+$; {\it squares}: $^4$HeD$^+$;
{\it green solid curve}: $^4$HeH$^+$ fit (Table~\ref{fits}); {\it blue dashed curve}: $^4$HeD$^+$ fit (Table~\ref{fits}).}
\label{HeH+_b}
\end{figure}

 HeH$^+$+H$\rightarrow$ H$_2^+$+He\\

\noindent which, together with H+H$^+$ radiative association, is the main formation pathway for the molecular hydrogen cation.
The equilibrium, permanent dipole moment of HeH$^+$ is approximately equal to 1.66 D (Pavanello et al.~2005).
Engel et al.~(2005)  computed  line lists for different HeH$^+$ isotopologues ($^3$HeH$^+$,$^4$HeH$^+$,
$^3$HeD$^+$,$^4$HeD$^+$), and data are available electronically\footnote{exomol - Molecular Line lists for Exoplanet Atmospheres, www.exomol.com} both for ro-vibrational levels and Einstein coefficients.
The cooling functions for all the isotopologues are shown in Figs. \ref{HeH+_a} and \ref{HeH+_b}.

\subsection{LiH}
LiH represents the primordial molecule with the largest permanent equilibrium
dipole moment ($\approx 5.88$~D, Hertzberg~1978).
Several studies have modelled in detail
lithium kinetics in the primordial Universe chemistry (e.g. Bougleux \& Galli~1997, Lepp et al.~2002),
also explicitly evaluating the cooling properties of lithium hydride due to
collisional processes.
The cooling properties have been described by GP98, where a fit of the collisional cooling function
in the low density limit was given.

$^7$LiH has a singlet $X \, ^1 \Sigma^+$ ground state with an experimental equilibrium distance
$r_e = 3.015$~a$_0$ and a dissociation energy 
$D_e = 20,287.7$~cm$^{-1}$ (Stwalley \& Zemke~1993).

In virtue of its small number of electrons LiH has been the object of
many quantum mechanical studies, among which we may mention
the pioneering one by \cite{Mulliken1936} and the 
recent ones by \cite{bande2010}, \cite{Cooper2009} and \cite{Gadea2009}.
These studies, however,
do not provide in a readily usable format
the potential energy curve and dipole moment curve we need 
to compute the cooling function. We therefore independently computed the required quantities.

\subsubsection{Computational detail}\label{section-LiH}
The potential energy and dipole moment curves have been calculated using the program
\cite{ref.molpro}, supplemented with the \cite{mrcc} package. 
Descriptions of the quantum chemical methods used, along with references to the
original papers, can be found, e.g., in the review by \cite{Lodi2010}.

The basis sets used for hydrogen belong to aug-cc-pV$n$Z, ($n = $ T, Q, 5)
family by \cite{Dunning1989}.
For lithium we used the recent aug-cc-pwC$n$Z(-DK) ($n=$ T, Q, 5) basis sets by \cite{prasher}.
Basis sets with the DK suffix were especially devised to be used with the 
relativistic {\sc dkh} hamiltonian (see below).
Computationally it would be entirely possible to use larger basis sets but, regrettably, no larger ones 
are available for lithium.

The restricted Hartree-Fock ({\sc rhf}) method applied to LiH wrongly dissociates to the ionic limit
Li$^+$(1$s^2$)+H$^{-}$(1$s^2$) 
instead of the correct neutral pathway Li(1$s^2$2$s^1$)+H(1$s^1$). 
This behaviour is a direct consequence of the {\sc rhf} model, which upon dissociation
restricts molecular fragments to be spin-singlets (see e.g. Lodi \& Tennyson~2010).
Because the dissociation limit is wrong, methods such as {\sc ccsd(t)}
which rely on the {\sc rhf} solution being a good approximation
to the exact wave function may be expected to
experience problems at long bond lengths.
In fact, the {\sc ccsd(t)} module in {\sc molpro} does not converge
for $r$ greater than about 6~a$_0$ (using the default settings). 
This may be slightly suprising because, as LiH has only 2 valence electron, method such as {\sc ccsd(t)}
are formally equivalent to full configuration interaction ({\sc fci}) in the valence space and should therefore be quite accurate.
It is the non-linearity of the equations to be solved, together with the poor starting guess, 
that result in these numerical problems.

We calculated for six geometries all-electron full configuration interaction ({\sc fci})
reference energies in the aug-cc-pwCQZ basis sets
using the {\sc ccsdtq} code of {\sc mrcc}.
Computation of each {\sc fci} point involves about $36 \times 10^6$
configuration state functions and took between 30h and 50h on a 4-core Intel Xeon 5160
workstation with 16~GiB of memory.
We compared these {\sc fci} values with several single-reference and multi-reference methods;
the results of the comparisons, not discussed here, indicate that a very good balance of
speed and accuracy is provided by the internally-contracted, multi-reference averaged coupled pair
functional ({\sc ic-acpf}) method based on a full-valence (2-electron, 5-orbital) complete active space
(8 reference configurations).
This method was therefore used for further calculations.
With respect to the {\sc fci} energies the {\sc ic-acpf} method predicts a dissociation energy
too low by about 6.0~cm$^{-1}$.
Computation of a {\sc ic-acpf} energy in the aug-cc-pCw5Z-DK basis set took about 4~min on the same Xeon 5160 machine.

To account for scalar-relativistic effects we considered the use of the traditional mass-velocity
one-electron Darwin operator ({\sc mvd1}) or, alternatively, of the fourth-order
Dirac-Kroll-Hess hamiltonian ({\sc dkh4}).
Relative energies computed by the two methods agree to better
than $\pm 0.05$~cm$^{-1}$. We chose the {\sc dkh4} approach.
Relativistic corrections are small, affecting relative energies only by  $\pm 2$~cm$^{-1}$. 
The contribution of relativistic corrections to the dissociation energy is $-0.5$~cm$^{-1}$.
Finally, we remark that in our tests differences in relative energies between DK-contracted and
aug-cc-pwC$n$Z basis sets are negligible, less than $\pm 0.05$~cm$^{-1}$ for
energies up to dissociation. 

The energy and dipole curves were computed for 300 uniformely-spaced grid points for
$r =1.00$~a$_0$ to $r = 16.00$~a$_0$ with spacing $0.05$ a$_0$.
The final energy values were obtained for each grid point
by basis-set extrapolation of the $n=$T, Q and 5 values;
following common practice (see, e.g., Lodi \& Tennyson~2010) the {\sc cas-scf} energies
were extrapolated using the functional form $E_n = E_\infty + A e^{-\alpha n}$ while
the differences between {\sc ic-acpf} and {\sc cas-scf} energies were extrapolated
using $E_n = E_\infty + A / n^3 $.

A minor complication arose because, for all basis sets employed,
the {\sc cas-scf} energies experience a discontinuity jump of
about 6~cm$^{-1}$ around $r=11$~a$_0$ if {\sc rhf}
orbitals are used as starting guess; as a result the {\sc ic-acpf} energies
also experience a discontinuity jump of about 1.2~cm$^{-1}$.
This problem is undoubtedly another consequence of the poor {\sc rhf} starting guess; it
 was circumvented by starting calculations at small bond lengths and then
concatenating subsequent calculations for increasing $r$ so that
the {\sc cas-scf} orbitals from the preceeding geometry were used
as starting guess.

\begin{figure}
\includegraphics[width=0.5\textwidth]{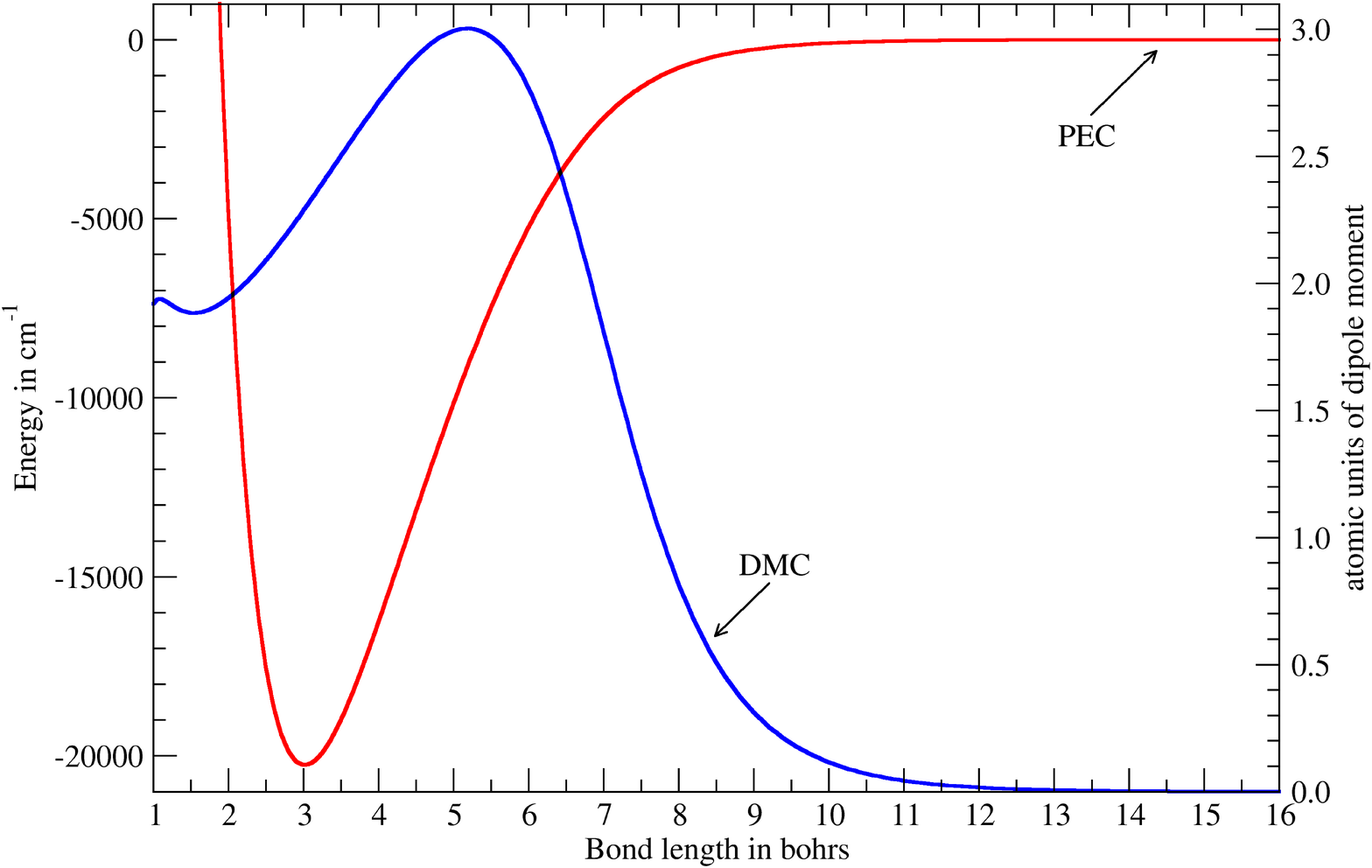}
\caption{LiH potential energy curve ({\sc pec}) and dipole moment curve ({\sc dmc}). The two curves use different scales for the $y$ axis; the left-hand side legend refers to the {\sc pec}, the right-hand side one to the {\sc dmc}.}\label{LiH-plot}
\end{figure}

The final potential energy curve has an equilibrium bond length $r_e = 3.014$~a$_0$ and a dissociation energy $D_e=20,294$~cm$^{-1}$; the dipole moment curve has an equilibrium dipole moment of 2.293~a.u., reaching
a maximum of 3.003~a.u. for $r = 5.180$~a$_0$.
We may assign to $D_e$ an uncertainty of the order of $\pm 15$~cm$^{-1}$, mainly due to basis set incompleteness.
For comparison, the non-relativistic `exact BO' value quoted by \cite{Cooper2009} is $D_e = 20,298.8$~cm$^{-1}$.
The {\sc pec} and the {\sc dmc} are plotted in Fig.~\ref{LiH-plot}.

\begin{figure}
\includegraphics[width=0.5\textwidth]{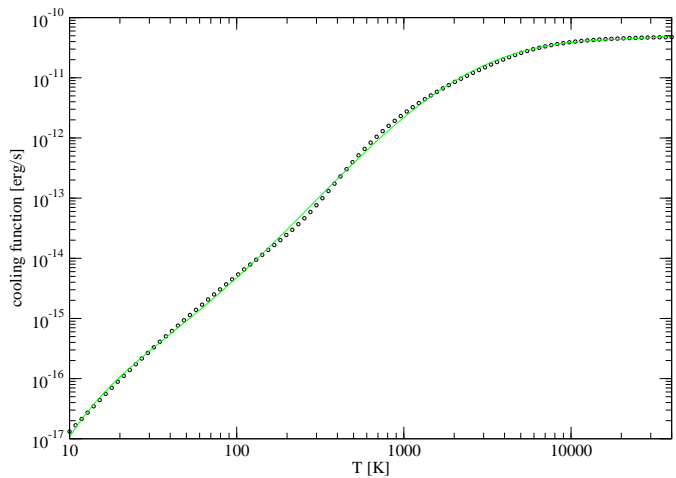}
\caption{LiH radiative cooling function.{\it Circles}: present calculation; {\it green solid curve}: fit given in Table~\ref{fits}.}
\label{LiH}
\end{figure}

\subsection{LiH$^+$}
LiH$^+$ represents the most abundant molecular species containing lithium at low redshift. Indeed, while in
the very beginning phases LiH is much more abundant than its cation, this behaviour reverses at $z\sim20$,
as predicted by Dalgarno \& Lepp~(1987) and subsequently found by kinetics
models (e.g. GP98, Bougleaux \& Galli~1997). The reason
of this behaviour is the presence of a residual 
ionization fraction which enhances one of the 
main formation channel, namely ion-atom radiative association.

So far LiH$^+$ has not been characterised experimentally.
The present calculations, discussed below, predict a very modest
dissociation energy $D_e$ of 1130.5~cm$^{-1}$, an equilibrium bond length
$r_e = 4.130$~a$_0$ and a permanent equilibrium dipole (measured in the
molecular centre-of-mass reference frame) of 0.2773~a.u. (0.7048~D).

Very specialised, high-accuracy methods are applicable to LiH$^+$ due to its small size.
Very recently \cite{Bubin2011} presented non-adiabatic, relativistic-corrected
values for the first five rotationless ($J=0$) vibrational states with a stated accuracy of better than
$0.1$~cm$^{-1}$.
Among other recent studies treating LiH$^+$ we may mention \cite{Gadea2009} and \cite{Magnier2004}.
However, again as in the case of LiH, because of the different focus these recent
studies do not provide in a usable-format both the
potential energy curve and the dipole moment curve we need to compute the cooling function.
We therefore independently computed the necessary quantities.

\subsubsection{Computational details}
Like the LiH case discussed in Section~\ref{section-LiH}
we used the aug-cc-pwC$n$Z-DK ($n = $ T, Q, 5) basis sets
and we accounted for scalar relativistic effects by specifying
the {\sc dkh4} hamiltonian. Calculations were done using \cite{ref.molpro}.

The restricted open-shell Hartree-Fock method ({\sc rohf}) applied to LiH$^+$
correctly dissociates to Li$^+$ + H.
Single-reference methods are therefore expected to fare very well, in contrast
to the LiH case discussed in Section~\ref{section-LiH}.

As a preliminary test we compared using the aug-cc-pwCTZ-DK basis set
38 full configuration interaction
({\sc fci}) relative energies in the range 3.00--11.00~a$_0$ with
energies computed by various other methods. 
Here is a brief summary of the comparison,
in the form {\sc method}/error: {\sc rohf}/98.9~cm$^{-1}$;
{\sc cisd}/1.8~cm$^{-1}$;
{\sc uccsd}/1.7~cm$^{-1}$; {\sc cisd+p}/1.4~cm$^{-1}$;
{\sc cisd+q}/0.6~cm$^{-1}$; {\sc uccsd(t)}/0.3~cm$^{-1}$.
The reported errors are one half of the
non-parallelity error, which is defined as the difference between the
maximum and the minimum of the modulus of the energy differences
between two methods (Li \& Paldus~1995).
As expected, {\sc uccsd(t)} produces extremely accurate results; also
note that this method is asymptotic to {\sc fci}
upon dissociation as it is size-extensive and exact for
two-electron systems.
We therefore used {\sc rhf-uccsd(t)} as implemented in \cite{ref.molpro} for further calculations.

Because LiH$^+$ is a charged species the value of the dipole moment depends on the choice of
the origin of the axes. As discussed by \cite{Bunker-Jensen}, the correct choice
of the origin is the molecular centre-of-mass. 
{\sc molpro} by default does set the origin at the centre-of-mass but uses
isotopically-averaged nuclear masses, which is not appropriate in our case.
We therefore specified the following masses: $m({\mathrm{Li}^+)}=7.015455$~u and
$m(\mathrm{H})=1.0078250321$~u. These are \emph{atomic} masses, to partially account
for the electron contribution to the centre-of-mass coordinates;
$m({\mathrm{Li}^+)}$ was obtained by subtracting one electron mass
to the atomic mass of Li.


With this choice of masses and taking the molecule to lie along the $x$ axis 
the coordinates of the nuclei for an inter-nuclear distance $r$ are:
$x(\mathrm{Li}^+) = 0.125613 r$ and $x(\mathrm{H}) = -0.874387 r$.
The dipole moment $M(r)$ is therefore asymtotic
to $M(r) \to 0.125613 r$ in this reference system. 

\begin{figure}
\includegraphics[width=0.5\textwidth]{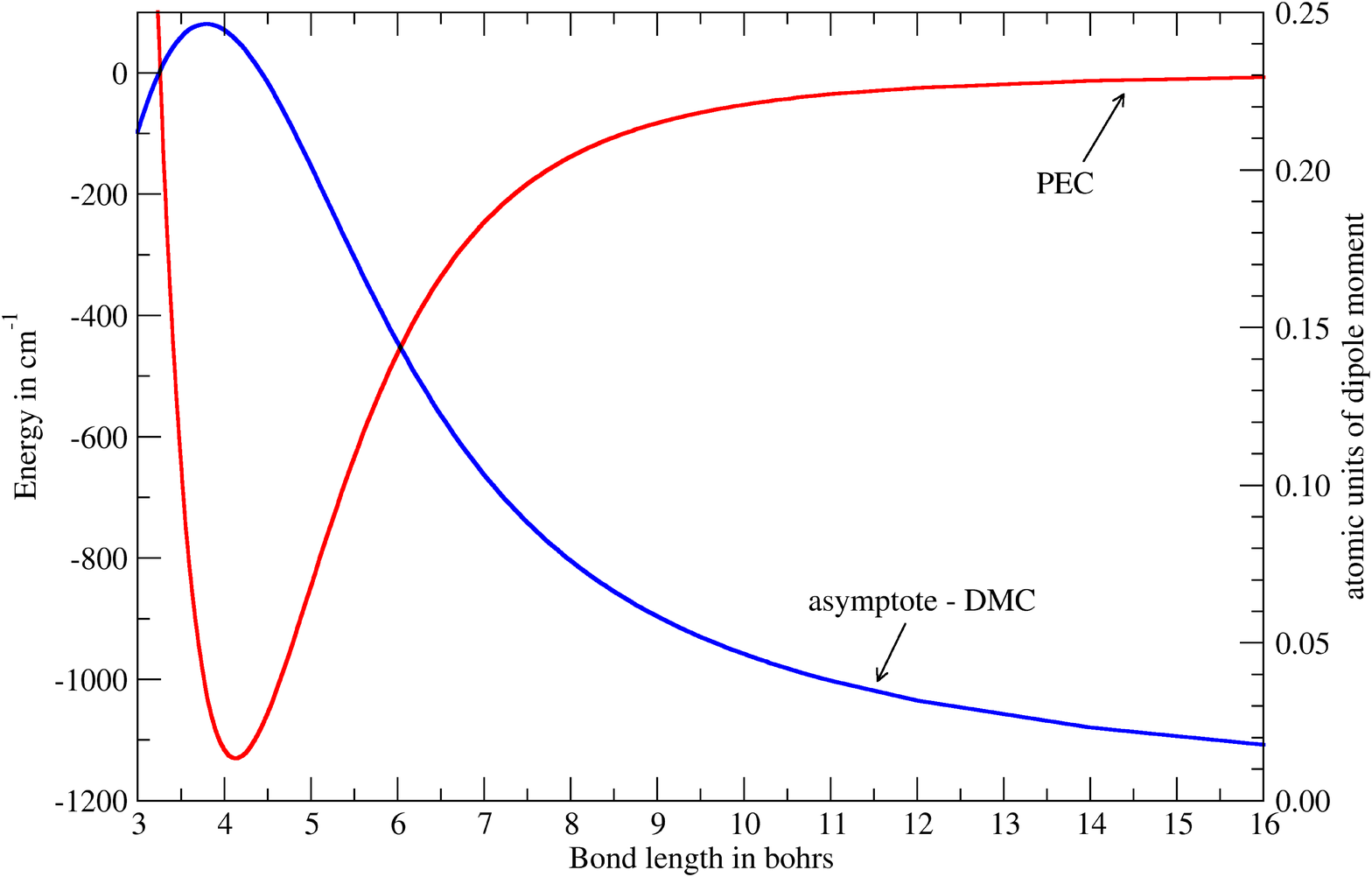}
\caption{LiH$^+$ potential energy curve ({\sc pec}) and the quantity $0.125613 r - M(r)$ (see text for details). The two curves use different scales for the $y$ axis; the left-hand side legend refers to the {\sc pec}, the right-hand side one to the {\sc dmc}.}\label{plot-LiH+}
\end{figure}

We present as figure~\ref{plot-LiH+} a plot of the {\sc pec} and of the
quantity $0.125613 r - M(r)$, namely the difference between the
asymptotic dipole and the actual one. We chose not to plot directly $M(r)$
as, because of the asymptotic form, it would very nearly appear as a straight line.

We report here below the rotationless ($J=0$) vibrational terms $E(\nu+1)-E(\nu)$ together
with the difference with the very accurate values by \cite{Bubin2011} for $\nu=0,1,2,3,4$:
354.59(0.44); 261.79(0.78); 170.00(0.59); 89.82(-0.07); 35.26(0.18).
Our computed spectroscopic values are therefore very accurate,
with errors of less than 1~cm$^{-1}$.

\begin{figure}
\includegraphics[width=0.5\textwidth]{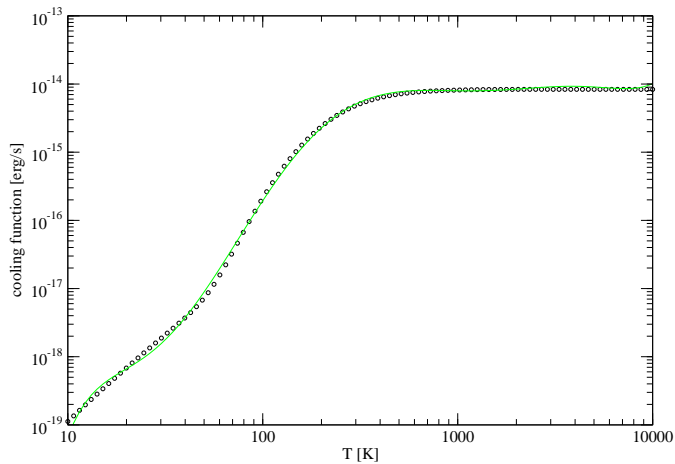}
\caption{LiH$^+$ radiative cooling function: {\it circles}: present calculation; {\it green solid line}: fit given in Table~\ref{fits}.}
\label{LiH+}
\end{figure}

\subsection{H$_3^+$}
H$_3^+$ is one of the most interesting molecular ion in astrophysics, especially for its role in the interstellar medium chemistry. Its cooling properties
 have been recently considered, both in planetary
 atmosphere conditions by \cite{b18bis} and in the primordial Universe by \cite{b8}. The former work gives
a fit for the radiative cooling function as a function of temperature based
on the {\it ab initio} line list of Neale et al.~(1996).

\subsection{H$_2$D$^+$}
In the deuterium chemistry of primordial Universe, H$_2$D$^+$ represents the most complex triatomic molecule usually introduced
 \citep{stancil,vol}. Recently, Sochi \& Tennyson (2010) have calculated a comprehensive line list of  H$_2$D$^+$ frequencies
 and transition probabilities. Table~\ref{fits} provides an improved fit for the radiative cooling function at low temperatures.

\begin{figure}
\includegraphics[width=0.5\textwidth]{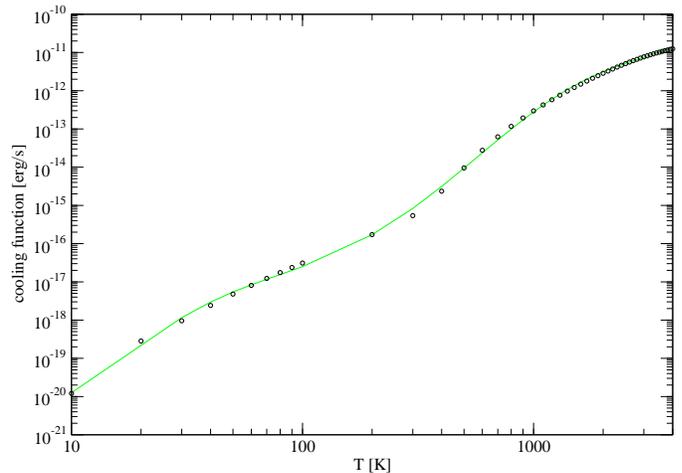}
\caption{H$_2$D$^+$ radiative cooling function: {\it circles}: calculation by Sochi \& Tennyson~2010; {\it green solid line}: fit given in Table~\ref{fits}.}
\label{LiH+_sochi}
\end{figure}

\section{Conclusions}\label{conclusioni}
In the present work, the cooling functions of the most abundant molecular species in the primordial Universe have been
studied; calculations of the radiative contributions have been taken into account, under the hypothesis of
local thermal equilibrium distributions of ro-vibrational levels.
Radiative cooling functions are shown for each molecule. Analytic fits to
each of these functions were obtained in the form:
\begin{equation}
\log_{10}~W =\sum_{n=0}^N a_{n}(\log_{10} T)^{n}.
\end{equation}
Coefficients for these fits, $a_{n}$,
are provided in Table~\ref{fits}. The fits are valid only in the
temperature range specified in the figures.
Potential energy and dipole moment curves calculated in the present work,
together with the computed transition wavenumbers and Einstein
coefficients,
can be downloaded from www.exomol.com.

\begin{table*}
 \centering
 \begin{minipage}{140mm}
  \caption{Radiative cooling function fits.}
  \begin{tabular}{@{}lllllllll@{}}
  \hline
   Molecule  &  & Coefficients&Molecule  &  & Coefficients&Molecule  &  & Coefficients\\
 \hline
HD &N=5 &$a_0=-55.5725$ &HD$^+$ &N=7&$a_0=-6.04917$& $^3$HeH$^+$ &N=7 &$a_0=-10.2807$ \\
  &&$a_1=56.649$&   &&$a_1=-60.0312$ &&&$a_1=-62.9415$ \\
  &&$a_2=-37.9102$&  &&$a_2=98.8361$  &&&$a_2=118.348$\\
  &&$a_3=12.698$&   &&$a_3=-77.5575$ &&&$a_3=-97.3721$\\
  &&$a_4=-2.02424$&  &&$a_4=33.4951$  &&&$a_4=42.5517$\\
  &&$a_5=0.122393$& &&$a_5=-8.07092$  &&&$a_5=-10.1946$\\
  &&&                &&$a_6=1.01514$  &&&$a_6=1.26335$\\
  &&&                &&$a_7=-0.0519287$ &&&$a_7=-0.0633433$\\
  &&&&&&&&\\
$^4$HeH$^+$ &N=7  &$a_0=-7.58736$&$^3$HeD$^+$ &N=7 &$a_0=26.2717$& $^4$HeD$^+$ &N=7 &$a_0=28.5384$ \\
  &&$a_1=-68.2966$&                 &&$a_1=-168.493$                              &&&$a_1=-172.458$ \\
  &&$a_2=122.847$&                  &&$a_2=247.288$                               &&&$a_2=249.811$\\
  &&$a_3=-99.444$&                  &&$a_3=-184.299$                               &&&$a_3=-184.786$\\
  &&$a_4=43.1409$&                  &&$a_4=77.0755$                               &&&$a_4=76.9479$\\
  &&$a_5=-10.3034$&                 &&$a_5=-18.2012$                              &&&$a_5=-18.1306$\\
  &&$a_6=1.27565$&                  &&$a_6=2.26293$                               &&&$a_6=2.25209$\\
  &&$a_7=-0.0639846$&               &&$a_7=-0.11512$                             &&&$a_7=-0.114559$\\
  &&&&&&&&\\
LiH &N=6  &$a_0=-31.894$ &LiH$^+$ &N=7  &$a_0=-23.5448$ &H$_2$D$^+$ &N=6  &$a_0=33.8462$ \\
  &&$a_1=34.3512$&   &&$a_1=9.93105$    &&&$a_1=-188.249$ \\
  &&$a_2=-31.0805$&  &&$a_2=-8.6467$    &&&$a_2=253.037$\\
  &&$a_3=14.9459$&   &&$a_3=2.13166$    &&&$a_3=-168.02$\\
  &&$a_4=-3.72318$&  &&$a_4=2.43072$    &&&$a_4=59.3597$\\
  &&$a_5=0.455555$&   &&$a_5=-1.69457$   &&&$a_5=-10.6334$\\
  &&$a_6=-0.0216129$&   &&$a_6=0.384871$   &&&$a_6=0.759029$\\
  &&&                &&$a_7=-0.0300114$ &&&\\
&&&&&&&&\\
H$_3^+$ &&fit by \cite{b18bis}&&&&&&\\
  \hline
\label{fits}
\end{tabular}
\end{minipage}
\end{table*}

\section*{Acknowledgments}

We thank Eveline Roueff and Herve Abgrall for making available their calculations on dipolar and quadrupolar emission
probabilities of HD. C.M.C. would also acknowledge Savino Longo for helpful discussion on collisional cooling functions, and MIUR
and Universit\`{a} degli Studi di Bari, that partially supported this work (\textquotedblleft fondi di Ateneo 2010\textquotedblright).

\bsp

\label{lastpage}

\end{document}